# Giant enhancement of anomalous Hall effect in Cr modulation-doped non-collinear antiferromagnetic $Mn_3Sn$ thin films

Xin Chen[1], Hang Xie[1], Qi Zhang[1,2], Ziyan Luo[3], Lei Shen[4,5], and Yihong Wu[1,6,*]

[1]*Department of Electrical and Computer Engineering, National University of Singapore, Singapore 117583*

[2]*Department of Electrical and Electronic Engineering, Southern University of Science and Technology, Xueyuan Rd 1088, Shenzhen 518055, China*

[3]*School of Physics and Electronics, Central South University, Changsha 410083, China*

[4]*Department of Mechanical Engineering, National University of Singapore, Singapore 117575*

[5]*Engineering Science Programme, National University of Singapore, Singapore 117575*

[6]*National University of Singapore (Chong Qing) Research Institute, Chongqing Liang Jiang New Area, Chongqing 401123, China*

E-mail: elewuyh@nus.edu.sg

We report on Cr doping effect in $Mn_3Sn$ polycrystalline films with both uniform and modulation doping. It is found that Cr doping with low concentration does not cause notable changes to the structural and magnetic properties of $Mn_3Sn$, but it significantly enhances the anomalous Hall conductivity, particularly for modulation-doped samples at low temperature. A Hall conductivity as high as 184.8 $\Omega^{-1}$ $cm^{-1}$ is obtained for modulation-doped samples at 50 K, in a sharp contrast to vanishingly small values for undoped samples at the same temperature. We attribute the enhancement to the change of Fermi level induced by Cr doping.

Recently $Mn_3X$ (X = Sn, Ge, Ga, Rh, Ir, Pt) based non-collinear antiferromagnets (AFMs) have attracted significant attention[1-10] because, unlike the widely studied collinear AFMs, these non-collinear AFMs exhibit large anomalous Hall,[1-4] anomalous Nernst[5] and magneto-optical Kerr[7] effects at room temperature and importantly, all these effects are accessible via either a moderate Oersted field or an effective spin torque field.[8-10] It has also been shown that these non-collinear AFMs host a wide range of exotic phenomena from magnetic Weyl fermions,[11,12] to ferroic ordering of cluster octupole moment,[13] spin polarized current[14] and magnetic spin Hall effect.[15] These exciting findings suggest that the chiral spin structure and the resultant Berry curvature affect profoundly the charge and spin transport properties and make the $Mn_3X$-based AFMs not only appealing for fundamental studies but also promising for spintronics applications. For the latter, it is important to prepare $Mn_3X$ thin films using commonly available deposition techniques such as sputtering, which has already been reported by several groups.[16-21] The initial results are encouraging as similar phenomena reported in bulk samples were also observed in thin films, albeit the size of the corresponding effect such as anomalous Hall effect (AHE) is typically much smaller than that of bulk samples, suggesting that further studies are required to improve the structural and physical properties of $Mn_3X$ thin films.

As the distribution of the non-zero Berry curvature and their relative positions to the Fermi level play an important role in determining the transport properties of $Mn_3X$-based AFMs, one of the possible pathways is to tune the Fermi level by chemical doping without significantly affecting the electronic band structures. In this context, we investigate Cr-doping effect on the structural, magnetic, and transport properties of $Mn_3Sn$, a representative $Mn_3X$-based AFM, via both uniform and modulation doping. It is found that both doping techniques lead to a large enhancement of AHE below room temperature compared to undoped samples, particularly for the modulation-doped samples. In addition to AHE, we also observe an enhancement of negative magnetoresistance (MR) when the external magnetic field and current are parallel.

All the samples are deposited on $SiO_2/Si$ substrates by magnetron sputtering at room temperature with a base pressure of $2 \times 10^{-8}$ Torr and a working pressure of $3 \times 10^{-3}$ Torr. Uniform doping is achieved through co-sputtering of $Mn_3Sn$ and Cr with the Cr composition controlled by the sputtering powers of both targets. As for modulation-doped samples, we form $[Mn_3Sn(t_1)/Cr(t_2)]_N$ multilayers and use the individual layer thickness to control the equivalent Cr composition; here $t_1$ and $t_2$ are the thicknesses of $Mn_3Sn$ and Cr in *nm* and $N$ is the repetition number. All the samples are post-annealed at 450℃ in vacuum with a base pressure of $2 \times 10^{-5}$ Torr for 1 hour to promote both crystallization and dopant diffusion. Crystalline structures of the samples are analyzed using X-ray diffraction (XRD) on a RIGAKU Smartlab system with Cu Kα radiation. Magnetic measurements are performed using SQUID (Superconducting Quantum Interference Device) magnetometry. For AHE and MR measurements, the samples are patterned into Hall bars with a lateral dimension of 150 μm (length) × 15 μm (width) using combined techniques of photolithography, sputtering and lift-off. The electrical measurements are performed in a Quantum Design's VersaLab system.

Figure 1 shows the XRD patterns of three groups of samples. The first group is a single layer $Mn_3Sn$ (label it as MS below) with a thickness of 72 nm. The second group consists of three $[Mn_3Sn(2.4)/Cr(t_2)]_{30}$ samples with $t_2 = 0.15, 0.3$ and 0.6 nm, respectively. In the third group are two co-sputtered samples deposited at the same $Mn_3Sn$ cathode power of 50 W, but with a different Cr cathode power of 10 W (MS50Cr10) and 30 W (MS50Cr30). The atomic compositions of Mn and Sn in $Mn_3Sn$ analyzed by X-ray photoelectron spectroscopy (XPS) are 0.79 and 0.21, respectively, which are clearly Mn-rich compared to the stoichiometric composition. On the other hand, the equivalent atomic compositions of $[Mn_3Sn(2.4)/Cr(0.3)]_{30}$ and $[Mn_3Sn(2.4)/Cr(0.6)]_{30}$ are Mn: 0.6 and 0.49, Sn: 0.19 and 0.15, and Cr: 0.21 and 0.36, respectively. The XRD peaks for $Mn_3Sn$ include $(10\bar{1}1)$, $(11\bar{2}0)$, $(20\bar{2}0)$, $(0002)$, $(20\bar{2}1)$, $(20\bar{2}2)$, $(22\bar{4}0)$, $(22\bar{4}2)$, and $(40\bar{4}1)$, indicating that it is polycrystalline with a mixture of crystallites with different orientations. The XRD peaks for $[Mn_3Sn(2.4)/Cr(t_2)]_{30}$ are more or less the same as that of $Mn_3Sn$ except that the $(0002)$ peak becomes smaller. On the other hand, the XRD patterns of the two co-sputtered samples are almost the same as that of the $Mn_3Sn$ sample including the peak positions. The lattice constant $a$ of $Mn_3Sn$ is calculated to be 5.673 Å, while those of Cr-doped samples are all around 5.666 Å, close to that of bulk $Mn_3Sn$ (~5.666 Å).[22] We have also conducted low angle X-ray reflectometry (XRR) measurements for $[Mn_3Sn(2.4)/Cr(t_2)]_{30}$ with $t_2 = 0.3, 0.6, 0.9, 1.2$ and 1.8 nm, but we didn't observe any peak due to the periodicity of the multilayer except for the $t_2 = 1.8$ nm sample, which exhibits a very weak peak. The above XRD and XRR results suggest that both the uniform and modulation-doped $Mn_3Sn$ exhibits similar crystalline structure as that of undoped samples, suggesting good solubility of Cr in $Mn_3Sn$ in both cases.

Figures 2 (a) – (d) show the *M-H* curves of these samples measured at 300 K by applying the field in different directions, namely, out-of-plane (red solid-line), in-plane 0° (green dotted-line), in-plane 45º (purple dashed-line) and in-plane 90º (black dash dotted-line). Here, we take one side edge of the sample as the reference for the in-plane field, *i.e.*, in-plane 0°. The out-of-plane saturation magnetizations ($M_s$) of [Mn$_3$Sn(2.4)/Cr(0.15)]$_{30}$, [Mn$_3$Sn(2.4)/Cr(0.6)]$_{30}$, and MS50Cr10 at 300 K are 7.28, 9.04, 7.03, and 8.07 emu/cm$^3$, respectively. These values are on the same order of the $M_s$ reported for polycrystalline Mn$_3$Sn films at 300 K.[20,23,24] The shape of both in-plane and out-of-plane *M-H* curves can be fitted well using the formula

$$M = \frac{2M_s}{\pi} \sum_{i=1}^{n} \alpha_i \, \mathrm{atan}[\beta_i(H \pm H_{ci})], \qquad (1)$$

where $M_s$ is the total saturation magnetization, $n$ is the number of a particular type of crystallites with a distinctive crystalline orientation, $\alpha_i$ is a proportional coefficient to denote the contribution from crystallites with different orientations with $\sum_{i=1}^{n} \alpha_i = 1$, $\beta_i$ indicates the projection of external field on the *c*-plane, and $H_{ci}$ is the coercivity when the field is applied in plane. The "$\pm$" sign indicates the forward and backward sweeping curves, respectively. We find that all the *M-H* curves can be fitted reasonably well with $n = 3$. The fitting results (Figures S1 and S2 in supplementary data) suggest that the combined contribution from $(11\bar{2}0)$, $(20\bar{2}0)$ and $(22\bar{4}0)$ planes to the out-of-plane magnetization of Mn$_3$Sn is about 35%, the other 50% is due to $(20\bar{2}1)$ and $(40\bar{4}1)$ planes, and the remaining 15% goes to $(10\bar{1}1)$ and $(20\bar{2}2)$ planes. For [Mn$_3$Sn(2.4)/Cr(0.15)]$_{30}$ the same ratios are 25%, 55% and 20%, respectively, indicating a slight change in the crystallite orientations. The results are consistent with the intensity ratio of XRD peaks. Figure 2(e) shows the temperature-dependence of $M_s$ for the four samples measured at 200 Oe during the warming up sweep after the sample is zero-field cooled to 50 K. As can be seen, the magnetization of Mn$_3$Sn decreases monotonically from 300 K to 250 K and becomes almost constant from 250 K to 50 K, indicating a gradual phase transition from the inverse triangular phase to the helical phase from 300 K to 250 K.[25] Below 250 K, a glassy ferromagnetic phase appears and co-exists with the helical phase at 50 K.[26] On the other hand, the magnetizations of [Mn$_3$Sn(2.4)/Cr(0.15)]$_{30}$ and MS50Cr10 both increase gradually from 300 K to 50 K and there is no inflection point, suggesting the absence phase transition and the inverse triangular spin structure is maintained down to 50 K. In contrast, the magnetization of [Mn$_3$Sn(2.4)/Cr(0.6)]$_{30}$ increases dramatically at around 210 K and reaches almost 9 times of that at 300 K when the temperature reaches 50 K. This is presumably caused by the transition from AFM to ferromagnetic or glassy ferromagnetic phase below 210 K, though further studies are required to determine the exact magnetic structure.

Next, we turn to the electrical transport properties. Figure 3(a) shows the schematic diagram of electrical measurements, and the Hall measurement results for the aforementioned four samples are given in Fig. 3(b). In general, the Hall resistivity of Mn$_3$Sn has three main contributions and can be described as

$$\rho_{xy} = R_0 B + R_s \mu_0 M + \rho_{xy}^{AF}, \qquad (2)$$

where $M$ is the magnetization, $R_0$ ($R_S$) is the ordinary (anomalous) Hall coefficient (both are positive), $B$ is the applied field, and $\mu_0$ is the magnetic permeability.[27] The last term, $\rho_{xy}^{AF}$, originates from the inverse triangular spin structure. As can be seen from Fig. 3(b), all the samples exhibit a negative AHE at 300 K, a hallmark of non-collinear AFM.[3,4,26] The zero-field $\rho_{xy}^{AF}$ for Mn$_3$Sn at 300 K is about -2.27 μΩ cm, comparable with or larger than the reported values for polycrystalline films.[16,17,24] When $T$ decreases to 250 K, the negative AHE decreases dramatically while a small positive component appears, indicating the transition from the AFM phase to the co-existence of helical and FM phases. This transition in AHE coincides with the transition observed in the *M-H* and *M-T* curves. When the temperature decreases further to 200 K, the negative component disappears and only a small positive component remains. This means that only the helical and FM phases exist in Mn$_3$Sn at 200 K. At 50 K, the AHE is vanishingly small due to the dominance of the spin glass phase.

In a sharp contrast, the temperature-dependence of AHE in [Mn$_3$Sn(2.4)/Cr(0.15)]$_{30}$ and [Mn$_3$Sn(2.4)/Cr(0.6)]$_{30}$ shows an opposite trend. In both samples, $\rho_{xy}^{AF}$ increases monotonically with decreasing temperature. In [Mn$_3$Sn(2.4)/Cr(0.15)]$_{30}$, $\rho_{xy}^{AF}$ at 300 K is 3.35 μΩ cm (hereafter we omit the negative sign), but it increases to 5.66 μΩ cm at 50 K, almost doubled from its room temperature value. Moreover, its room temperature value is also 47.6% higher than that of Mn$_3$Sn with almost the same thickness. On the other hand, $\rho_{xy}^{AF}$ of [Mn$_3$Sn(2.4)/Cr(0.6)]$_{30}$ increases from 1.48 μΩ cm at 300 K to 2.52 μΩ cm at 50 K, which changes by 70%, albeit the absolute value is much smaller than that of [Mn$_3$Sn(2.4)/Cr(0.15)]$_{30}$. In contrast, $\rho_{xy}^{AF}$ of MS50Cr10 increases from 3.02 μΩ cm at 300 K to 3.8 μΩ cm at 200 K and then decreases to 2.42 μΩ cm at 50 K. In addition to these four samples, we have also varied the Cr thickness in the multilayer samples and sputtering power of the co-sputtered samples systematically. The temperature-dependence of $\rho_{xy}^{AF}$ for all the samples are summarized in Fig. 3(c). The general trends are summarized as follows. In the case of [Mn$_3$Sn(t$_1$)/Cr(t$_2$)]$_{30}$ multilayers, when t$_1$ is fixed at 2.4 nm, t$_2$ in the range of 0.13 – 0.3 nm results in a significant enhancement of AHE throughout the temperature range investigated, *i.e.*, 50 – 300 K, as compared to Mn$_3$Sn. In fact, $\rho_{xy}^{AF}$ for [Mn$_3$Sn(2.4)/Cr(0.6)]$_{30}$ keeps increasing slowly down to 10 K. When t$_2$ is fixed at 0.15 nm, a reduction of t$_1$ from 2.4 nm to 2 nm results in a non-monotonic temperature-dependence of AHE similar to that of co-sputtered samples, and a further decrease of t$_1$ to 1 nm leads to a vanishingly small AHE. On the other hand, for co-sputtered samples, when the sputtering power for Mn$_3$Sn is fixed at 50 W, the samples deposited at a Cr sputtering power of 10 and 30 W show a similar non-monotonic temperature-dependence of AHE, though $\rho_{xy}^{AF}$ for the sample deposited at 10 W is much larger than that of the sample deposited at 30 W, 3.02 μΩ cm versus 1.79 μΩ cm, at 300 K. These results suggest that there is a difference between the co-sputtered and multilayer samples even though the equivalent Cr composition is almost the same. The multilayer samples are presumably comprised of alternative layers of heavily doped and lightly or non-doped Mn$_3$Sn. The heavily doped region helps to stabilize the AFM phase at low temperature and the lightly doped region leads to an enhancement of AHE. Figure 3(d) shows the temperature-dependence of zero-field anomalous Hall conductivity (AHC), $\sigma_{xy} = \frac{-\rho_{xy}}{\rho_{xx}^2}$, where $\rho_{xx}$ is the longitudinal resistivity at $H = 0$. At 300 K, $\sigma_{xy}$ of [Mn$_3$Sn(2.4)/Cr(0.15)]$_{30}$, [Mn$_3$Sn(2.4)/Cr(0.6)]$_{30}$ and MS50Cr10 are 57.8 Ω$^{-1}$ cm$^{-1}$, 52.6 Ω$^{-1}$ cm$^{-1}$ and 41.6 Ω$^{-1}$ cm$^{-1}$, respectively. When the temperature decreases to 50 K, it reaches 184.8 Ω$^{-1}$ cm$^{-1}$, 149.9 Ω$^{-1}$ cm$^{-1}$ and 109.2 Ω$^{-1}$ cm$^{-1}$, respectively. As a comparison, $\sigma_{xy}$ for Mn$_3$Sn is 39.5 Ω$^{-1}$ cm$^{-1}$ at 300 K and nearly 0 at 50 K. It is apparent that Cr-doping significantly enhances AHC, especially at low temperature. In addition to the large AHE, we have also observed large negative magnetoresistance (NMR) with *H//I* in Cr-doped samples at low temperature (Figure S3 in supplementary data), around -0.15% for [Mn$_3$Sn(2.4)/Cr(0.15)]$_{30}$ at 50 K. The NMR, which is due to chiral anomaly,[11,27] is reported to be inversely proportional to the energy spacing between the Weyl nodes and Fermi

level.[27)] Therefore, both the AHE and NMR results can be accounted for by the doping-induced narrowing of energy spacing between the Weyl nodes and Fermi level.

In summary, the Cr doping effect in $Mn_3Sn$ polycrystalline films has been investigated. Both uniform and modulation doping with low Cr concentration do not cause notable changes to structural and magnetic properties of $Mn_3Sn$, but it significantly enhances the anomalous Hall effect, and the enhancement is particularly significant for modulation-doped samples at low temperature with the magnitude much larger than that of its bulk counterpart. The enhancement is presumably caused by the change of Fermi level induced by Cr doping. Compared to uniform doping, modulation doping has the advantage that the dopant distribution is more regular in the thickness direction which helps to suppress doping-induced disorder, as manifested in the large enhancement of AHE, especially at low temperature. Our results may open new opportunities for both device applications and fundamental studies of $Mn_3Sn$ at low temperature.

**Acknowledgments**

This work is supported by Ministry of Education, Singapore under its Tier 2 Grants (grant no. MOE2017-T2-2-011 and MOE2018-T2-1-076). We thank Profs Shufeng Zhang, Ryotaro Arita, and Jingsheng Chen for helpful discussions.

## Figure Captions

**Fig. 1.** XRD spectra for $Mn_3Sn$ (72), $[Mn_3Sn(2.4)/Cr(t_2)]_{30}$ and co-sputtered samples MS50Cr30 and MS50Cr10.

**Fig. 2.** *M-H* curves measured at 300 K with the field applied out-of-plane (red solid-line), in-plane $0°$ (green dotted-line), in-plane $45^o$ (purple dashed-line) and in-plane $90^o$ (black dash dotted-line) for (a) $Mn_3Sn$, (b) $[Mn_3Sn(2.4)/Cr(0.15)]_{30}$, (c) $[Mn_3Sn(2.4)/Cr(0.6)]_{30}$ and (d) MS50Cr10. (e) *M-T* curves from 50 K to 300 K under a 200 Oe out-of-plane magnetic field.

**Fig. 3.** (a) Schematic of Hall bar and electrical transport measurement. (b) Magnetic field-dependence of anomalous Hall resistivity for $Mn_3Sn$, $[Mn_3Sn(2.4)/Cr(t_2)]_{30}$ and MS50Cr10 at different temperatures. (c) Temperature-dependence of zero-field Hall resistivity for different samples. (d) Temperature-dependence of zero-field AHC for $Mn_3Sn$, $[Mn_3Sn(2.4)/Cr(t_2)]_{30}$ and MS50Cr10.

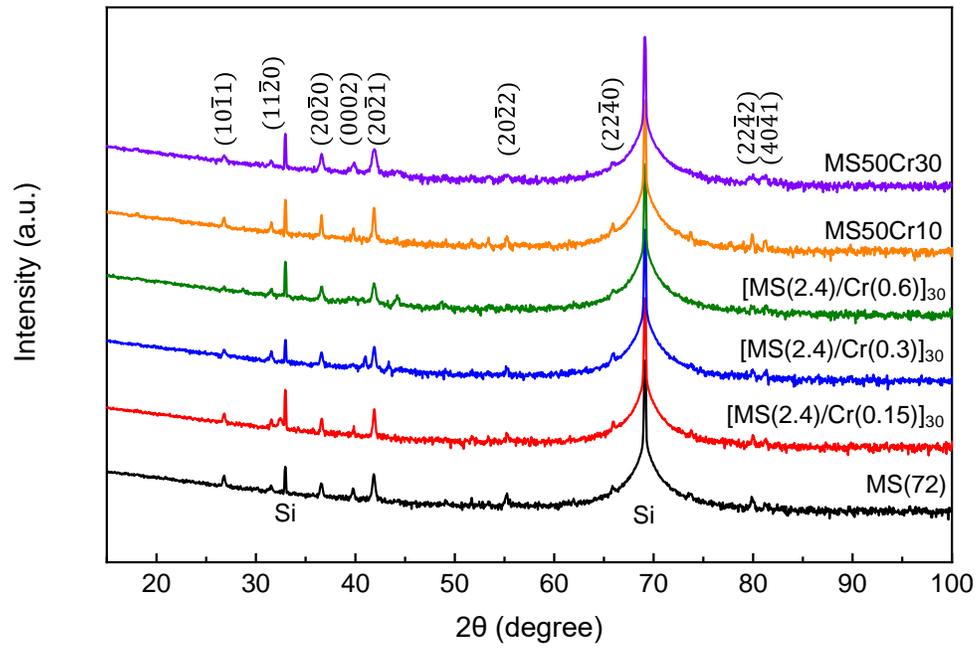

Figure 1

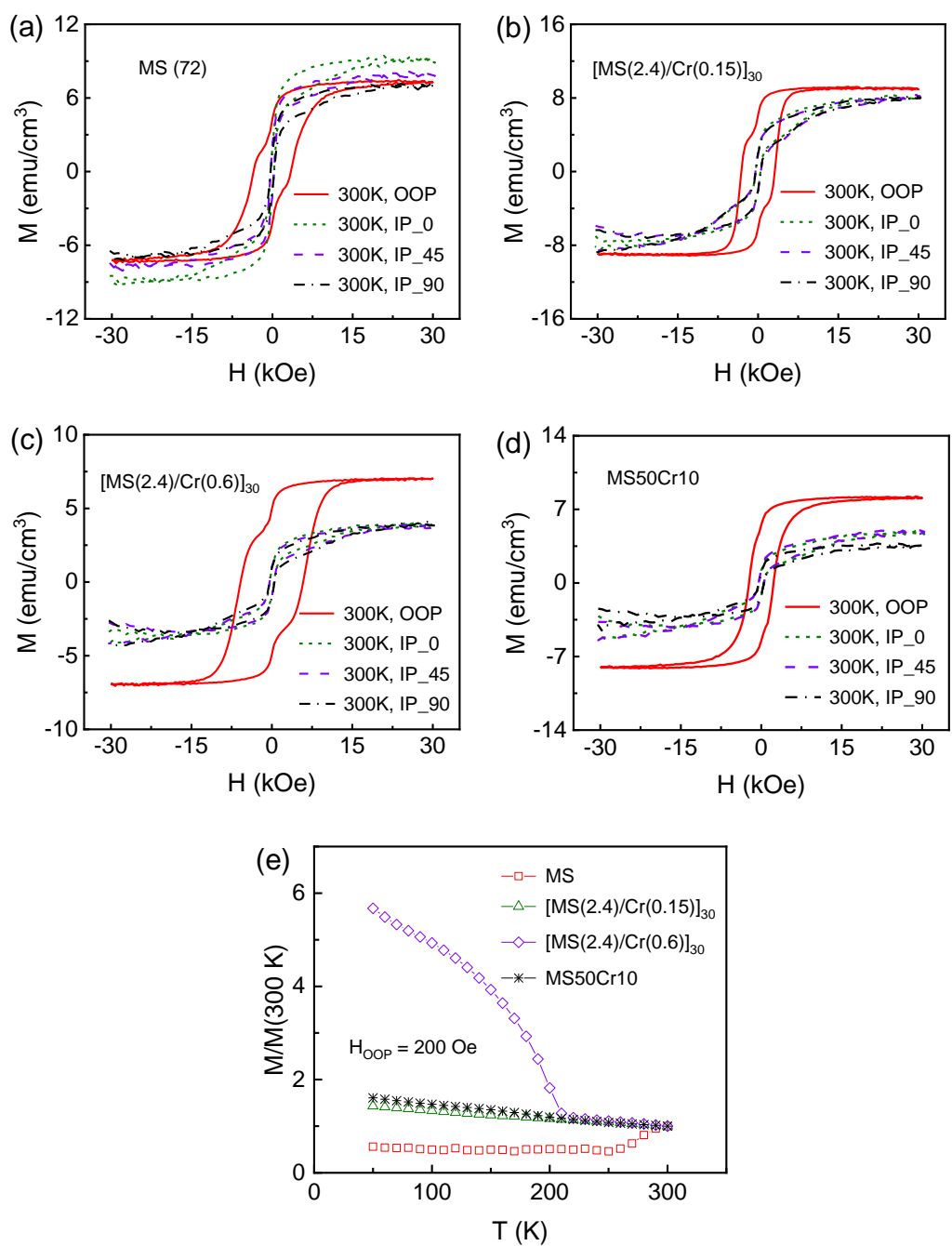

Figure 2

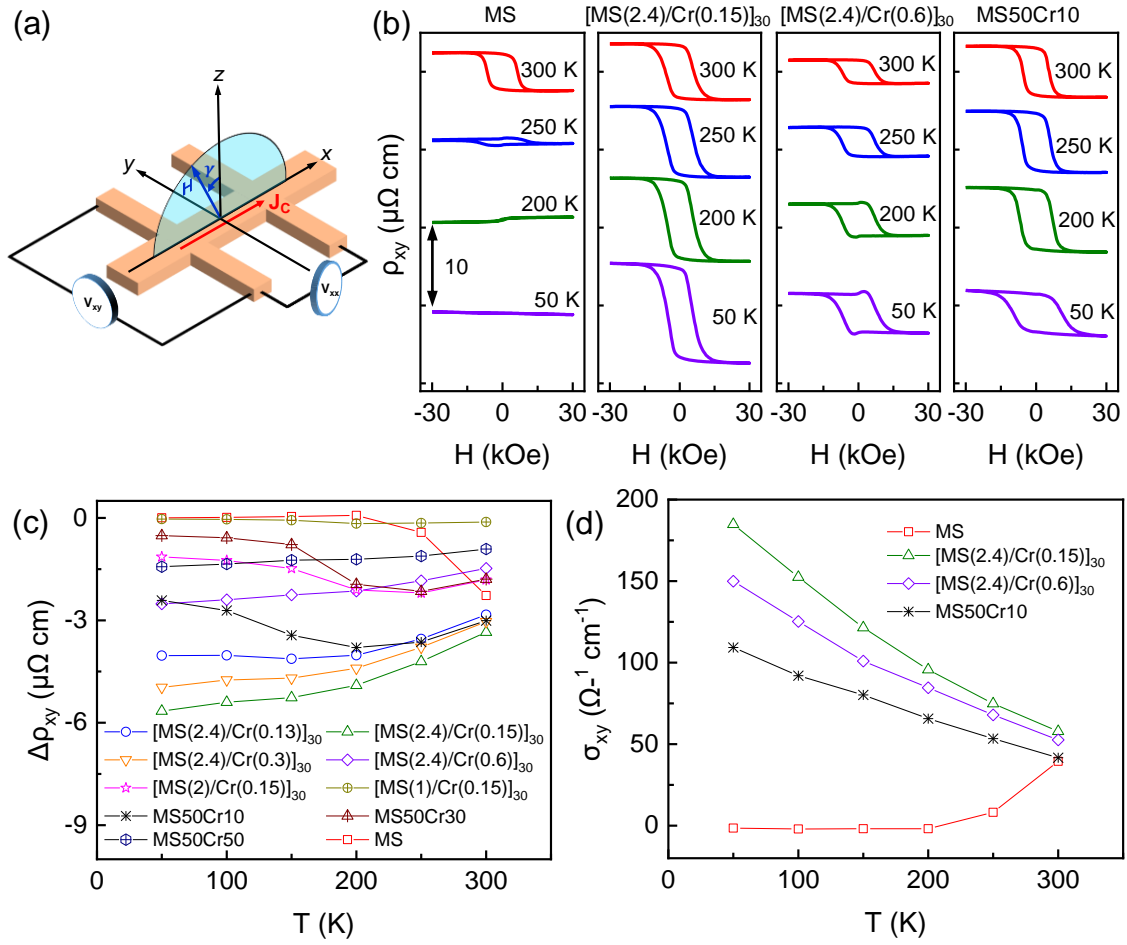

Figure 3